\documentclass[aps,prb,twocolumn,amsfonts,amssymb,amsmath,floats,floatfix,showpacs,preprintnumbers,superscriptaddress]{revtex4-2}

\usepackage{amsmath}
\usepackage{graphicx}
\usepackage{natbib}
\usepackage{ulem} 
\usepackage{color,soul}

\DeclareGraphicsExtensions{.pdf,.eps,.png,.jpg,.mps}

\usepackage{epstopdf}

\usepackage{color}
\usepackage{dcolumn}
\usepackage{bm}
\usepackage{hyperref}
\usepackage{relsize}
\def \Mgb{MgB$_{2}$} 

\hypersetup{
colorlinks=true,
urlcolor= blue,
citecolor=blue,
linkcolor= blue,
bookmarksopen=false,
}

\begin{document}

\title{Selective electron-phonon coupling strength \\from nonequilibrium optical spectroscopy: The case of MgB$_2$}

\author{S. Mor}
\email{selene.mor@unicatt.it}
\affiliation{I-LAMP (Interdisciplinary Laboratories for Advanced Materials Physics), Universit\`{a} Cattolica del Sacro Cuore, via della Garzetta 48, Brescia I-25133, Italy}
\affiliation{Department of Physics, Universit\`{a} Cattolica del Sacro Cuore, via della Garzetta 48, Brescia I-25133, Italy}
\author{F. Boschini}
\affiliation{Institut National de la Recherche Scientifique -- \'{E}nergie Mat\'{e}riaux T\'{e}l\'{e}communications Varennes QC J3X 1S2 Canada}
\affiliation{Department of Physics {\rm {\&}} Astronomy, University of British Columbia, Vancouver, British Columbia V6T\,1Z1, Canada}
\affiliation{Quantum Matter Institute, University of British Columbia, Vancouver, British Columbia V6T 1Z4, Canada}
\author{E. Razzoli}
\author{M. Zonno}
\author{M. Michiardi}
\author{G. Levy}
\affiliation{Department of Physics {\rm {\&}} Astronomy, University of British Columbia, Vancouver, British Columbia V6T\,1Z1, Canada}
\affiliation{Quantum Matter Institute, University of British Columbia, Vancouver, British Columbia V6T 1Z4, Canada}
\author{N.D. Zhigadlo}
\affiliation{Laboratory for Solid State Physics, ETH Zurich, CH-8093 Zurich, Switzerland}
\affiliation{CrystMat Company, CH-8037 Zurich, Switzerland}
\author{P.C. Canfield}
\affiliation{Ames Laboratory, U. S. DOE, Iowa State University, Ames, Iowa 50011, USA}
\affiliation{Department of Physics and Astronomy, Iowa State University, Ames, Iowa 50011, USA}
\author{G. Cerullo}
\affiliation{Dipartimento	di	Fisica,	Politecnico	di	Milano,	Piazza L. da	Vinci	32,	I-20133	Milano,	Italy}
\affiliation{IFN-CNR, Piazza L. da Vinci 32, I-20133 Milano, Italy}
\author{A. Damascelli}
\affiliation{Department of Physics {\rm {\&}} Astronomy, University of British Columbia, Vancouver, British Columbia V6T\,1Z1, Canada}
\affiliation{Quantum Matter Institute, University of British Columbia, Vancouver, British Columbia V6T 1Z4, Canada}
\author{C. Giannetti}
\email{claudio.giannetti@unicatt.it}
\affiliation{I-LAMP (Interdisciplinary Laboratories for Advanced Materials Physics), Universit\`{a} Cattolica del Sacro Cuore, via della Garzetta 48, Brescia I-25133, Italy}
\affiliation{Department of Physics, Universit\`{a} Cattolica del Sacro Cuore, via della Garzetta 48, Brescia I-25133, Italy}
\affiliation{CNR-INO (National Institute of Optics), via Branze 45, Brescia I-25133, Italy}
\author{S. Dal Conte}
\email{stefano.dalconte@polimi.it}
\affiliation{Dipartimento	di	Fisica,	Politecnico	di	Milano,	Piazza L. da	Vinci	32,	I-20133	Milano,	Italy}

\begin{abstract}
The coupling between quasiparticles and bosonic excitations rules the energy transfer pathways in condensed matter systems. The possibility of inferring the strength of specific coupling channels from their characteristic time scales measured in nonequilibrium experiments is still an open question. Here, we investigate \Mgb, in which conventional superconductivity at temperatures as high as 39 K is mediated by the strong coupling between the conduction electrons and the E$_{2g}$ phonon mode. By means of broadband time-resolved optical spectroscopy, we show that
this selective electron-phonon coupling dictates the nonequilibrium optical response of \Mgb\, at early times ($<$100 fs) after photoexcitation. Furthermore, based on an effective temperature model analysis, we estimate its contribution to the total electron-boson coupling function extracted from complementary equilibrium spectroscopy approaches, namely optical reflectivity and ARPES. The coupling strength with the E$_{2g}$ phonon modes is thus estimated to be $\lambda \simeq$ 0.56, which is approximately half of the total coupling constant, in agreement with ab-initio calculations from the literature.
As a benchmark, broadband time-resolved optical spectroscopy is performed also on the isostructural and non-superconducting compound AlB$_2$, showing that the nonequilibrium optical response relaxes on a slower time scale due to the lack of strongly-coupled phonon modes. Our findings demonstrate the possibility to resolve and quantify selective electron-phonon coupling from nonequilibrium optical spectroscopy.
\end{abstract}

\pacs{xxx}

\maketitle
\section{Introduction}
The dynamics of electronic quasiparticles (QPs) in solids is profoundly affected by the coupling with bosonic excitations, such as phonons and spin fluctuations. These interactions give rise to a renormalization of the electronic self-energy, which takes the form of a frequency-dependent effective mass and a scattering probability \cite{Carbotte1990}. In the paradigmatic case of superconducting materials, the electron-boson coupling strength determines the gap size and the critical temperature of the system, and therefore it is considered the crucial parameter to characterize the superconducting instability \cite{Carbotte2011,Scalapino2012,Marsiglio2020}. Angle-resolved photoemission spectroscopy (ARPES) and optical spectroscopy (OS) are the go-to techniques to extract the frequency-dependent electron-boson coupling and calculate the total coupling strength. ARPES is a surface-sensitive technique which directly
accesses the renormalized electronic band dispersion, and allows one to reconstruct the electronic self-energy with frequency and momentum resolution \cite{Damascelli2004}. However, complete frequency-momentum mapping is time consuming and strongly relies on sample surface preparation and achievable experimental resolution. On the other hand, OS is bulk sensitive and has access to the frequency-dependent optical constants, from which it is possible to obtain, via suitable modeling, the optical self-energy and the frequency-dependent QP scattering rate \cite{Basov2002,Basov2011}. Although less affected by sample preparation and experimental conditions, OS provides momentum-integrated information that requires complex model-dependent analysis to extract the QP scattering properties. In both cases, the measured self-energy is related to the electron-boson coupling function through an integral relation that depends on a temperature and frequency dependent kernel \cite{Carbotte2011,Basov2011,Giannetti2017}. Thus, the extraction of electron-boson coupling is an indirect operation subject to intrinsic uncertainties. Even more, both equilibrium approaches do not allow one to disentangle selective couplings with different modes that may coexist on the same energy scale \cite{Carbotte2011}.   

An alternative strategy is based on the observation that the electron-boson coupling function determines the electron relaxation time scales when the electronic population is driven out of equilibrium, for example via the use of ultrashort laser pulses \cite{Allen1987,Giannetti2017}.  In the last years, time-resolved spectroscopy methods have been widely employed to address how photoexcited charge carriers release their excess energy to different bosonic degrees of freedom~\cite{Giannetti2017,Na2020, Mor2022, Boschini2024}. In particular, time-resolved optical spectroscopy (tr-OS) has proven the ability to resolve the hierarchy of QP relaxation channels in the presence of selective coupling to specific bosonic modes with different coupling strengths. For instance, tr-OS of cuprates~\cite{DalConte2012,DalConte2015} has revealed a composite QP relaxation pathway which involves the energy exchange with spin fluctuations on a time scale of $\simeq$20 fs, with strongly coupled phonons (SCPs) within $\simeq$100 fs, and with the rest of the lattice vibrational modes within tens of picoseconds (ps). Despite the great progress of tr-OS, the quantification of the electron-boson coupling strength is based on the applicability of effective temperature models, which assume to separately resolve the thermalization of the electronic and each bosonic degree of freedom~\cite{Giannetti2017}. Actually, the reliability of such models, as well as the possibility of
quantifying the electron-boson
coupling function via time-resolved spectroscopic techniques, is still at
the center of an open debate \cite{Fann1992, Sentef2013,Giannetti2017,Riffe2024}.

In this work, we focus on a paradigmatic Bardeen-Cooper-Schrieffer (BCS) superconductor, magnesium diboride (\Mgb), which is characterized by a very strong and selective coupling of electrons with E$_{2g}$ phonon modes at $\simeq$70~meV that is responsible for the superconducting pairing~\cite{Nagamatsu2001}. We demonstrate that these SCPs impact on the QP relaxation dynamics measured by tr-OS, and that the strong electron-phonon coupling (EPC) directly reflects in an additional decay channel of the transient optical response. We benchmark these findings against aluminum diboride (AlB$_2$), an isostructural non-superconducting compound which lacks any selective electron-phonon coupling. The combination of tr-OS data with equilibrium OS and complementary ARPES allows us to demonstrate the capability of the nonequilibrium approach to discern the coupling with a subset of SCPs from that with the rest of the phonon modes, as well as to quantify the relevant coupling strengths. 

The structure of the work is the following. First, we determine the total Eliashberg spectral function $\alpha^2F(\mathbf{k},\omega)$ (also referred to as glue function) in MgB$_2$ by applying a simplified histogram model to equilibrium OS and ARPES data. These techniques consistently highlight a strong EPC at about $\simeq$70 meV. Then, by measuring the transient reflectivity variation $\delta R(t)/R_{eq}$ over a broad near-infrared energy range and with high (sub-20-fs) temporal resolution, we find that the QP relaxation evolves along two decay channels: (1) with a fraction of E$_{2g}$ SCPs on a fast time scale, i.e. $<$100 fs; (2) with the remaining fraction of weakly-coupled phonon modes on a slower time scale, i.e. $>$1 ps. Specifically, starting from the total glue function determined by equilibrium OS, we successfully reproduce the time-domain dynamics with an effective temperature model and provide an estimate of the coupling constant, $\lambda_{SCP}$, between the electronic QPs and the subset of E$_{2g}$ SCPs. Finally, we corroborate our results by a comparative study of AlB$_2$. Our findings not only provide a consistent evaluation of the EPC strength in \Mgb, but also provide evidence for a population of 'hot' E$_{2g}$ phonons which has been predicted to originate from the strong and selective EPC\cite{Novko2020}.    

\section{Selective electron-phonon coupling in $\mathrm{M\lowercase{g}B}$$_2$}
In the quest for increasing the critical temperature $T_c$ of superconducting materials, strong coupling of electrons to a selected phonon branch is predicted to play a major role compared to systems characterized by an isotropic EPC~\cite{McMillan1968,Nakhmedov1996,Devereaux2004}. This is particularly significant in the case of \Mgb, which in 2001 was discovered as the phonon-mediated superconductor with the highest critical temperature $T_c$ of 39\,K~\cite{Nagamatsu2001}. This  $T_c$ is twice as large as what estimated for an isotropic system with the same average EPC strength~\cite{Choi2002,Choi2002a,Eiguren2008,Margine2013}. 
In \Mgb, the EPC is concentrated in the bulk $\sigma$ electronic bands strongly coupled to the E$_{2g}$ optical phonon mode at $\sim$70~meV, corresponding to in-plane stretching of the B-B bonds~\cite{Kong2001,An2001,Yildirim2001,Bohnen2001}. The application of the Eliashberg theory has shown that this strong and selective coupling accounts for the high $T_c$ of the system, as opposed to the weak EPC of the $\pi$ bands\cite{Souma2003, Kong2001}. Recently, ab-initio theory predicted that the EPC selectivity of \Mgb\,should produce 'hot' phonons upon photoexcitation of the metallic state, suggesting the possibility of sustaining non-thermal states for unexpectedly long time~\cite{Novko2020,CAPPELLUTI2022,Cappelluti2022_2}. However, until now, experimental evidence of 'hot' phonons in \Mgb\, was reported only indirectly by tr-OS measurements through the onset of an unconventional behavior at the plasma frequency at ca. 6.5~eV~\cite{Baldini2017}. 
In contrast to \Mgb, the coupling with in-plane boron vibrations is strongly suppressed in AlB$_2 $, impeding the onset of superconductivity \cite{Bohnen2001} and the build-up of a 'hot' phonon population\cite{Novko2020}. Thus, \Mgb\, and AlB$_2$ are ideal platforms for investigating the role of SCPs on the QP relaxation dynamics, and addressing the capability of tr-OS to quantify the strength of a selective EPC.

\section{Quasiparticle self-energy and electron-phonon coupling}
\label{sec:theory}
At a given equilibrium temperature $T$, the electron-phonon interaction determines 
the renormalization of both the electronic band dispersion and lifetime, which is accounted for by the complex electronic self-energy $\Sigma(\mathbf{k},\omega,T)$, with $\mathbf{k}$ and $\omega$ being the electron wavevector and frequency, respectively. The real part of $\Sigma(\mathbf{k},\omega,T)$ defines the change of effective mass caused by the electron-phonon interaction that leads to a deviation (kink) of the electronic band dispersion. The inverse of the imaginary part, Im$^{-1}\Sigma(\mathbf{k},\omega,T)$, provides the electron lifetime. After integration over all possible electron-phonon scattering wavevectors, $\Sigma(\mathbf{k},\omega,T)$ can be calculated as the convolution integral between the Eliashberg spectral function $\alpha^2F(\mathbf{k},\Omega)$ and a kernel function $L(\omega,\Omega,T)$ \cite{Giannetti2017}. The former corresponds to the electron-phonon coupling function with $\alpha^2$ the squared matrix element of the interaction, the latter accounts for the Fermi distribution of electrons and the Bose distribution of phonons. Thus, under the assumption of a constant density of states, the self-energy reads:
\begin{equation}
\Sigma(\mathbf{k},\omega,T)=\int_0^{\infty}\alpha^2F(\mathbf{k},\Omega) L(\omega,\Omega,T) d\Omega.
\label{eq:SE}
\end{equation} 
By integrating over the entire Brillouin zone, it is possible to calculate the momentum-averaged Eliashberg function, $\alpha^2\bar{F}(\Omega)$, from which the total EPC strength $\lambda$ is obtained as:
\begin{equation}
\label{lambda}
\lambda= 2\int^{\omega}_{0} d\Omega \frac{\alpha^2\bar{F}(\Omega)}{\Omega}.
\end{equation} \\
\textcolor{black}{For strongly-coupled superconductors, the critical temperature $T_c$ is linked to the EPC strength $\lambda$ and the frequency of the coupled phonons $\Omega$ through the McMillan's formula~\cite{McMillan1968}:
\begin{equation}
    T_c = 0.83\Omega\text{ exp}\Bigg[\frac{-1.04(1-\lambda)}{\lambda-\mu^*(1+0.62\lambda)}\Bigg],
    \label{eq:Tc}
    \end{equation}
    with $\mu^*$ the screened Coulomb pseudopotential. In the case of multiple couplings, the partial contribution of each coupling channel to $T_c$ enables the identification of the pairing mechanism, as showcased for cuprates in Ref.~\onlinecite{DalConte2012}.}

ARPES directly probes the single-particle spectral function $A(\mathbf{k},\omega,T)$ which is related to the self energy through~\cite{Damascelli2004}:
\begin{widetext}
\begin{equation}
A(\mathbf{k},\omega,T)=-\frac{1}{\pi}\frac{\mathrm{Im}\Sigma(\mathbf{k},\omega,T)}{[\omega-\epsilon^0_{\mathbf{k}}-\mathrm{Re}\Sigma(\mathbf{k},\omega,T)]^2+[\mathrm{Im}\Sigma(\mathbf{k},\omega,T)+i\gamma_{\mathrm{imp}}]^2},
\end{equation}
\end{widetext}
where $\epsilon^0_{\mathbf{k}}$ is the bare electronic band dispersion (i.e. neglecting electron-phonon interaction) and $\gamma_{\mathrm{imp}}$ an intrinsic decay rate that accounts for the electron scattering by impurities. Thus, ARPES allows to retrieve the band-specific $\alpha^2F(\mathbf{k},\Omega)$ through the inversion of Eq. (\ref{eq:SE}), which is used in Eq. (\ref{lambda}) to calculate the band-selective EPC strength~\cite{Carbotte2011,Damascelli2004}.

Equilibrium OS measures the complex optical conductivity, $\sigma(\omega)$, which is determined by the $\mathbf{k}$-space integration of all possible electron-hole excitations. If we restrict to the conduction electrons, we can model the optical response by taking advantage of the extended Drude (ED) model, which is characterized by a frequency-dependent QP scattering rate \cite{Basov2011,Giannetti2017}. In this framework, the so-called optical memory function, $M(\omega,T)$, can be obtained through the Kubo formula in the Migdal approximation (omitting vertex corrections) as:
\begin{widetext}
\begin{equation}
\label{M}
M(\omega,T)=\omega\left\{\int_{-\infty}^{+\infty}\frac{f(\xi,T)-f(\xi+\omega,T)}{\omega+\Sigma^*(\xi,T)-\Sigma(\xi+\omega,T)+i\gamma_{\mathrm{imp}}}d\xi\right\}^{-1}-\omega,
\end{equation}
\end{widetext}
where $f$ is the Fermi–Dirac distribution. In this case, the momentum-integrated $\alpha^2\bar{F}(\omega)$ is obtained by inverting Eq. (\ref{eq:SE}), using e.g. maximum entropy methods \cite{Schachinger2006}.

As introduced in Ref. \citenum{Allen1987}, the Eliashberg spectral function also regulates the energy flow between electrons and phonons when the respective temperatures, $T_{el}$ and $T_{ph}$, are transiently decoupled. The functional describing the energy transfer rate between the two populations is given by:
\begin{widetext}
\begin{equation}
\label{G}
g(\alpha^2\bar{F},T_{el},T_{ph})= \frac{6\gamma_{el}}{\pi\hbar k_{\mathrm{B}}^{2}}\int_{0}^{\infty}\alpha^2\bar{F}(\omega)\omega^2\left[n\left(\omega,T_{ph}\right)-n\left(\omega,T_{el}\right)\right]d\omega,
\end{equation}
\end{widetext}
where $n$ is the Bose distribution and $\gamma_{el}$ the electronic specific heat. Therefore, under the assumption of an effective thermalization of electrons and phonons, time-resolved experiments allow to extract $\alpha^2\bar{F}(\omega)$ from the relaxation dynamics of 'hot' electrons generated by excitation with a short laser pulse. We note that, according to the Matthiessen's rule, the selective coupling with a subset of phonon modes will manifest itself as an additional relaxation channel, whose time constant depends on the relevant EPC strength. In this case, the total glue function can be expressed as the sum of all contributions due to the couplings with specific phonon modes (indexed by $i$), i.e. $\alpha^2\bar{F}$ = $\sum_i\alpha^2\bar{F}_i$. Accordingly, Eq. (\ref{G}) decouples in distinct functionals from which the specific EPC constants can be obtained. 
\textcolor{black}{Finally, we note that when the electronic system couples to only one phonon mode, the Eliashberg spectral function simply reduces to a delta function centered at the frequency of the coupled phonon. As a result, the self-energy and thus the scattering rate become frequency-independent, and the ED model reduces to the standard Drude model of simple metals. Here, given the energy-dependence of the electron-phonon coupling of \Mgb, we rather adopt the ED model to study the optical response in and out-of equilibrium.}

\section{Electron-phonon coupling by equilibrium optical spectroscopy}
\label{sec:OS}
\begin{figure}[t]
\centerline{\includegraphics[width=85mm]{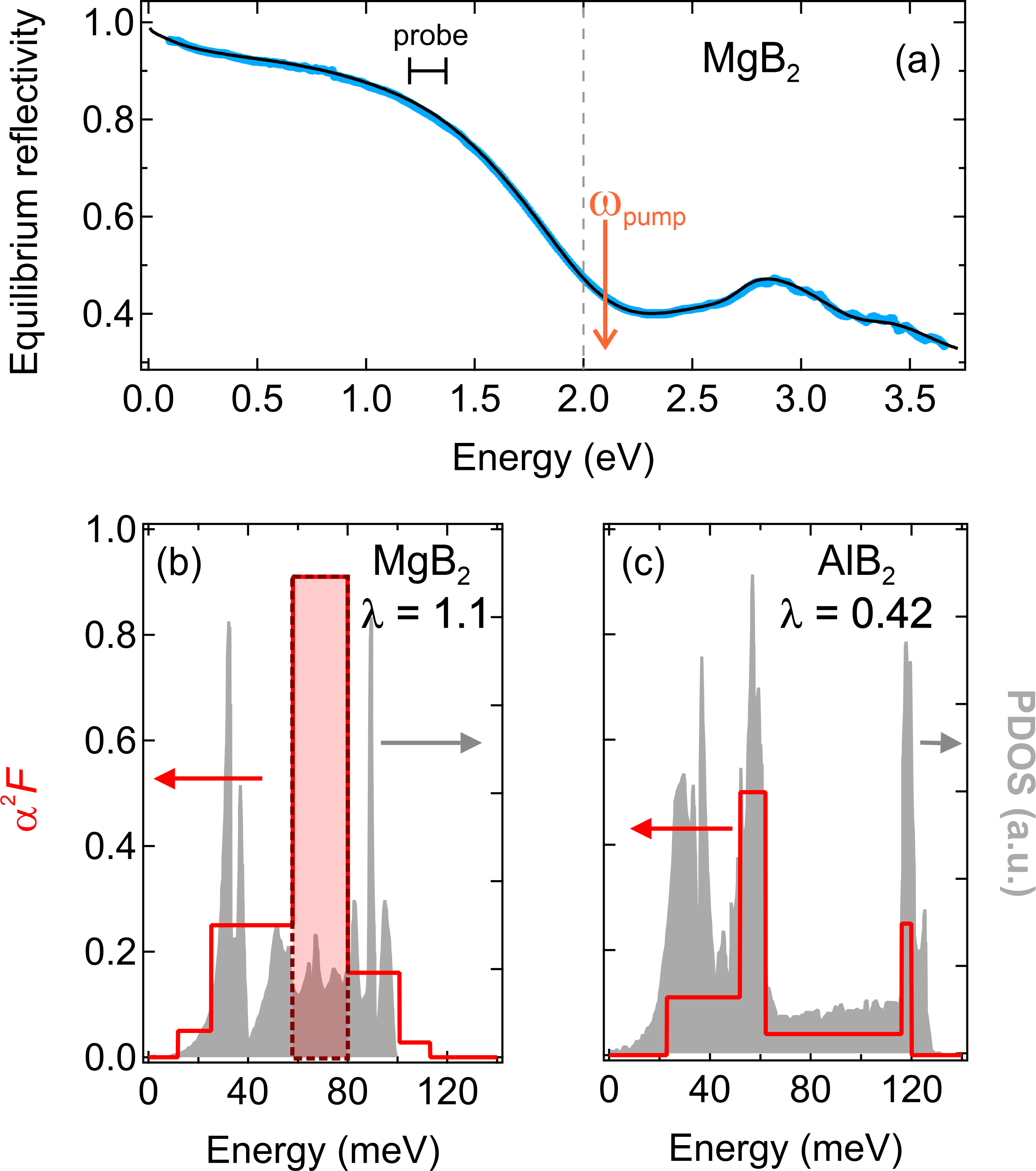}}
\caption{(a) Equilibrium reflectivity of \Mgb\,measured through spectroscopic ellipsometry (light blue line), and ED model fit to the data (black line). The black line is the fit to the data with an ED model and a sum of Lorentz oscillators which are located in the visible region above the plasma edge. The black horizontal bar indicates the photon energy region of the probe pulse in the tr-OS measurements, and the orange arrow the pump photon energy. The energy of the plasma edge is marked by the vertical dashed line.  (b) Momentum-integrated Eliashberg function $\alpha^2\bar{F}(\omega)$ (red histograms) of MgB$_2$ used to fit the data reported in (a), and calculated phonon density of states (PDOS, gray shade) taken from Ref.~\onlinecite{Kong2001} in arbitrary units. (c) Momentum-integrated Eliashberg function (red histograms) of AlB$_2$ used to approximate the calculated one from Ref.~\onlinecite{Bohnen2001}, and calculated PDOS (gray shade) also taken from Ref.~\onlinecite{Bohnen2001}.}
\label{Fig:Fig1}
\end{figure}

Fig. \ref{Fig:Fig1}(a) reports the \Mgb\, equilibrium reflectivity at $T$~=~300~K along the alternating layers of B and Mg atoms (light blue line, measurement taken from Ref.~\onlinecite{Guritanu2006}). It shows a broad low-energy structure with a plasma edge at about 2~eV and an isolated peak at 2.8~eV. 

The equilibrium reflectivity is related to the complex dielectric function $\epsilon(\omega)$ by:
\begin{equation}
\label{R}
R_{eq} = \left|\frac{1-\sqrt{\epsilon(\omega)}}{1+\sqrt{\epsilon(\omega)}}\right|^2,
\end{equation}
with
\begin{equation}
\label{epsilon}
\epsilon(\omega) = 1 + i\frac{4\pi\sigma(\omega)}{\omega},
\end{equation}
where $\sigma(\omega)$ is the complex optical conductivity. 
The low-energy part of the reflectivity spectrum is modeled by the ED optical conductivity $\sigma_{ED}(\omega)$ given by\cite{Heumen2009}:
\begin{equation}
\label{sigma}
\sigma_{ED}(\omega,T) = \frac{i}{4\pi}\frac{\omega^2_p}{\omega+M(\omega,T)}
\end{equation}
where $\hbar \omega_p\simeq$2 eV is the energy of the plasma edge and $M(\omega,T)$ is the optical memory function defined in Eq. (\ref{M}). By fitting the combination of Eqs. \ref{R}, \ref{epsilon} and \ref{sigma} to the experimental data, it is thus possible to extract the memory function that encodes the electron-boson scattering information. Following the successful approach previously adopted \cite{DalConte2012,Giannetti2017}, $\alpha^2\bar{F}(\omega)$ is modeled as a sum of histograms, whose heights and widths are used as fit parameters. The number of histograms is taken as the smallest one necessary to obtain a stable fit. Without expecting to reproduce the fine details of the EPC function (see Section I of the Supplemental Material for details~\cite{Supplementary}), this model enables successful fitting of the equilibrium reflectivity spectrum of \Mgb\,(black line on top of the data in Fig. \ref{Fig:Fig1}(a)). More importantly, it provides a robust estimation of the main electron-boson coupling features.

The retrieved $\alpha^2\bar{F}(\omega)$ is reported as red histograms in Fig.\ref{Fig:Fig1}(b). We note that the main contribution is centered at $\omega_{0}\sim$70 meV, which corresponds to the energy of the E$_{2g}$ B-B optical phonons. 
By using Eq. \ref{lambda}, we obtain a rather large value of the momentum-integrated EPC strength, $\lambda(\mathrm{MgB_{2}}) = 1.1\,\pm0.1$, in agreement with \textcolor{black}{experiment} reports in the literature~\cite{Dolgov2003,Carrington2003,Hwang2014} and consistent with the $T_c$ of the metal to superconductor phase transition \textcolor{black}{estimated using the McMillan's equation}. \textcolor{black}{Interestingly, ab-initio calculations typically report values between 0.6 and 0.7~\cite{Novko2020, Volpato2025}. While this discrepancy partly stems from the challenge of appropriately accounting for the strong anisotropy of the EPC in \Mgb~\cite{Liu2001}, it is well known that DFT tends to underestimate EPC in the presence of strong electron–boson interactions~\cite{Yin2013}.  Going back to Fig.\ref{Fig:Fig1}(b),} we also stress that the phonon density of states (PDOS) (gray shade) is almost constant over the whole energy range~\cite{Kong2001}. However, the EPC constant solely obtained by the contribution of the histogram between 60 and 80 meV (see the red area in Fig. \ref{Fig:Fig1} (b)) accounts for $\sim$50$\%$ of the total coupling. This result suggests the predominant and strong coupling of electrons to specific optical phonon modes  at about 70 meV, although a quantitative measure of $\alpha^2\bar{F}_{\mathrm{SCP}}(\omega)$ i.e. the contribution to the spectral function specific to SCP, is prevented by the overlap with the coupling to other phonon modes in the same energy range.   

For comparison, Fig. \ref{Fig:Fig1}(c) reports, on the left axis, the $\alpha^2\bar{F}(\omega)$ (red histograms) of AlB$_2$ that we obtained upon approximation of the calculated momentum-integrated Eliashberg function from Ref.~\onlinecite{Bohnen2001} with the same number of histograms used for \Mgb. On the right axis, we show the PDOS (gray shade), taken from Ref.~\onlinecite{Bohnen2001}. The absence of any discrepancy between the PDOS and the $\alpha^2\bar{F}(\omega)$ spectrum indicates the lack of the strong coupling with specific phonon modes that characterizes MgB$_2$. \textcolor{black}{In particular, we note that the peak in $\alpha^2\bar{F}(\omega)$ of AlB$_2$ at approximately 50 meV corresponds to a large PDOS rather than a stronger EPC with specific phonon modes at this energy. As a result, we do not either expect a population of hot phonons to build up in AlB$_2$ under out-of-equilibrium conditions~\cite{Cappelluti2022_2}. Rather, 
} 
we obtain a much lower EPC strength, $\lambda({\mathrm{AlB_{2}}}) = 0.42\,\pm\,0.1$, compatible with the lack of superconductivity in this isostructural compound~\cite{Nagamatsu2001}. 

\section{Selectivity of electron-phonon coupling by ARPES}
\begin{figure}[t]
\centerline{\includegraphics[width=85mm]{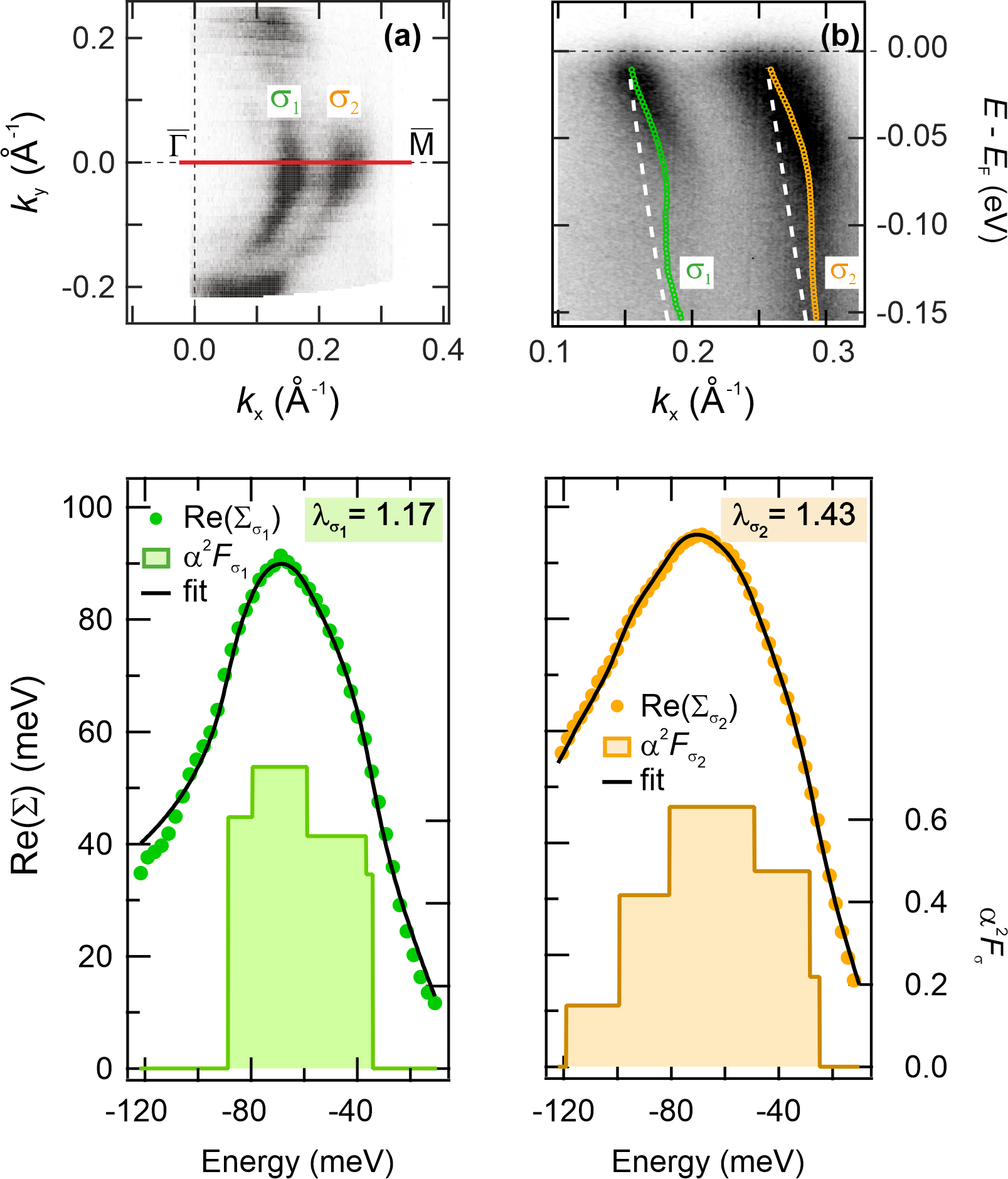}}
\caption{(a) ARPES mapping of the \Mgb\, Fermi surface. (b) ARPES dispersion of the $\sigma$ electronic bands along the $\Gamma-M$ direction; green and orange markers highlight the maxima of the momentum distribution curves (MDCs), obtained by fitting the MDCs for all binding energies (see main text for details).  The dashed lines indicate the non-interacting band dispersion. (c) - (d) Real part of the self energy (markers) obtained from the ARPES data (see main text), together with the best fit according to Eq. \ref{eq:SE} (black lines) and relevant glue function (histograms), for the two $\sigma$ bands.}
\label{Fig:Fig2}
\end{figure}
ARPES allows to estimate the EPC constant at a specific electron wavevector $\mathbf{k}$ by analysing the renormalization of the electronic band dispersion, also referred to as kink~\cite{Damascelli2003, Damascelli2004}. In the case of \Mgb, we are interested in the EPC constants of the $\sigma_1$ and $\sigma_2$ electronic bands which selectively couple with the E$_{2g}$ phonon mode. As introduced in Section \ref{sec:theory}, the renormalization of the electronic band dispersion depends on $\mathrm{Re}\Sigma(\mathbf{k},\omega,T)$, which is in turn determined by $\alpha^2F(\mathbf{k},\Omega)$.

Figure\,\ref{Fig:Fig2}(a) shows the Fermi surface mapping (20~meV-integration window about the Fermi energy $E_{\text{F}}$) of a MgB$_2$ single crystal measured at 10\,K. Prior to the measurement, the sample was oriented via Laue diffraction and cleaved in vacuum ($<$5$\cdot$10$^{-11}$\,Torr). Electrons were photoemitted by s-polarized 6.2~eV photons, and detected by a hemispherical electron analyzer (SPECS Phoibos 150) with overall momentum and energy resolutions of $<$0.003\,$\text{\AA}^{-1}$ and 17 meV. We resolve the separation between the two $\sigma$ electronic bands, which is maximal along the $\Gamma-M$ direction (red line), in agreement with previous laser-based~\cite{Mou2015,Mou2015a} as well as soft x-ray~\cite{Sassa2015} ARPES studies. The nonuniform distribution of the photoemission intensity is ascribed to the interplay between the p$_x$--p$_y$ orbital symmetries and the polarization of the incoming light, in addition to the effect of photoemission matrix elements~\cite{Damascelli2004}. The ARPES map of the $\sigma$ electronic bands along the $\Gamma-M$ direction is displayed in Fig.\,\ref{Fig:Fig2}(b). Both bands exhibit a significant kink in their energy-momentum dispersion at binding energy of approximately 70~meV with respect to $E_{\text{F}}$, in good agreement with recent ARPES measurements~\cite{Mou2015}. As this energy matches that of the E$_{2g}$ phonon mode, the kink points towards a strong EPC between the $\sigma$ electronic bands and the B-B stretching mode.

The renormalization of the electronic band dispersion is quantitatively analyzed through fitting of the momentum distribution curves (MDCs) at all energies with the sum of two Lorentzian functions accounting for the $\sigma_1$ and $\sigma_2$ electronic bands. The positions in momentum of the two Lorentzian peaks are shown by the green and orange markers in Fig.\,\ref{Fig:Fig2}(b). The white dashed lines are the bare electronic bands $\epsilon^0_{\mathbf{k}}$ with parabolic dispersion. In Fig. \ref{Fig:Fig2} (c) and (d), we present the deviation of the renormalized band dispersion from the bare band dispersion, i.e. $\omega_{\mathbf{k}}-\epsilon^0_{\mathbf{k}}$, which provides the real part of the self-energy for both $\sigma$ bands. Also shown is the best fit (black lines) obtained by Eq. (\ref{eq:SE}), together with the determined EPC function for the respective bands (histograms). Consistent with the analysis of the OS data, we adopted a simple model in which $\alpha^2F(\mathbf{k},\Omega)$ is approximated by a sum of histograms, whose parameters are determined by the data fitting. For both $\sigma$ bands, we obtain an EPC function peaking at $\sim$70 meV, which reflects the strong coupling of those bands to the B-B optical phonon mode. Moreover, we determine the total coupling for $\sigma_1$ and $\sigma_2$ to be 
$1.17\pm0.05$ and $1.43\pm0.05$, respectively. \textcolor{black}{We further test that these values can vary by approximately 20$\%$ when choosing a linear bare band dispersion, as reported e.g. in Refs. \citenum{Mou2015}, \citenum{Mou2015a}.} Overall, we find a good agreement with the momentum-integrated $\lambda$ obtained from the equilibrium OS measurement, as well as with calculations and experiments from the literature~\cite{Kong2001,Golubov2002,Mazin2002,Mazin2002a,Carrington2003,Eiguren2008,Mou2015,Dolgov2003,Hwang2014}. 
The difference between the two values agrees with previous results from the \textcolor{black}{literature }\cite{Liu2001,Mou2015,Mou2015a}. Analogous to the results obtained by OS, the ARPES-based self-energy reconstruction does not allow to isolate and quantify the specific contribution of the E$_{2g}$ mode to the total coupling since different phonon modes can contribute to the EPC strength in the same energy range.

\section{Selective electron-phonon coupling captured by tr-optical spectroscopy}
In contrast to equilibrium OS and ARPES, a strong EPC with specific phonon modes can be isolated and quantified in nonequilibrium conditions, since it gives rise to an additional relaxation channel that can be observed in the time-resolved optical response. To this end, it is important to link the measured transient variation of the optical properties, such as the reflectivity, to the actual effective electronic and bosonic temperatures of the system. As explained in the following, this approach is particularly appropriate when reflectivity below the plasma edge of a metallic system is measured. In the ED model, the reflectivity for $\omega<\omega_p$ 
can be approximated as:
\begin{equation}
 \label{Reflectivity}
 R(\omega,T)\simeq1-2\frac{\Gamma_{\mathrm{ED}}(\omega,T)}{\omega_p}
\end{equation}
where $\Gamma_{\mathrm{ED}}(\omega,T)$ = $\tau_{\mathrm{ED}}^{-1}(\omega,T)$ = $\mathrm{Im} M(\omega,T)$ is the frequency-dependent scattering rate. In a time-resolved experiment, the pump pulse can drive a transient reflectivity variation via two main mechanisms: i) an instantaneous increase of the electronic temperature decoupled from the bosonic bath, which \textcolor{black}{ produces a transient spectroscopic effect similar to that of} a transient increase of $\omega_p$ \cite{DalConte2012,DalConte2015,Giannetti2017} \textcolor{black}{(see Section IV of the Supplemental Material for details~\cite{Supplementary})}; ii) an increase of the bosonic temperature, which leads to an increase of the electron-boson scattering rate accounted for by $\Gamma_{\mathrm{ED}}$ \cite{DalConte2012,DalConte2015,Giannetti2017}. As a result, the transient reflectivity variation below $\omega_p$ is positive and described by $\delta R\simeq(2\Gamma_{\mathrm{ED}}/\omega^2_p)\delta \omega_p$ in the case of heating of the electron population decoupled from the bosonic bath, whereas it is negative and described by $\delta R\sim-(2/\omega_p)\delta \Gamma_{\mathrm{ED}}$, when the Bose-Einstein distribution of the bosonic excitations is heated by the pump pulse, thus leading to an increased scattering rate. 

\begin{figure*}
\centering
\includegraphics[width=1\textwidth]{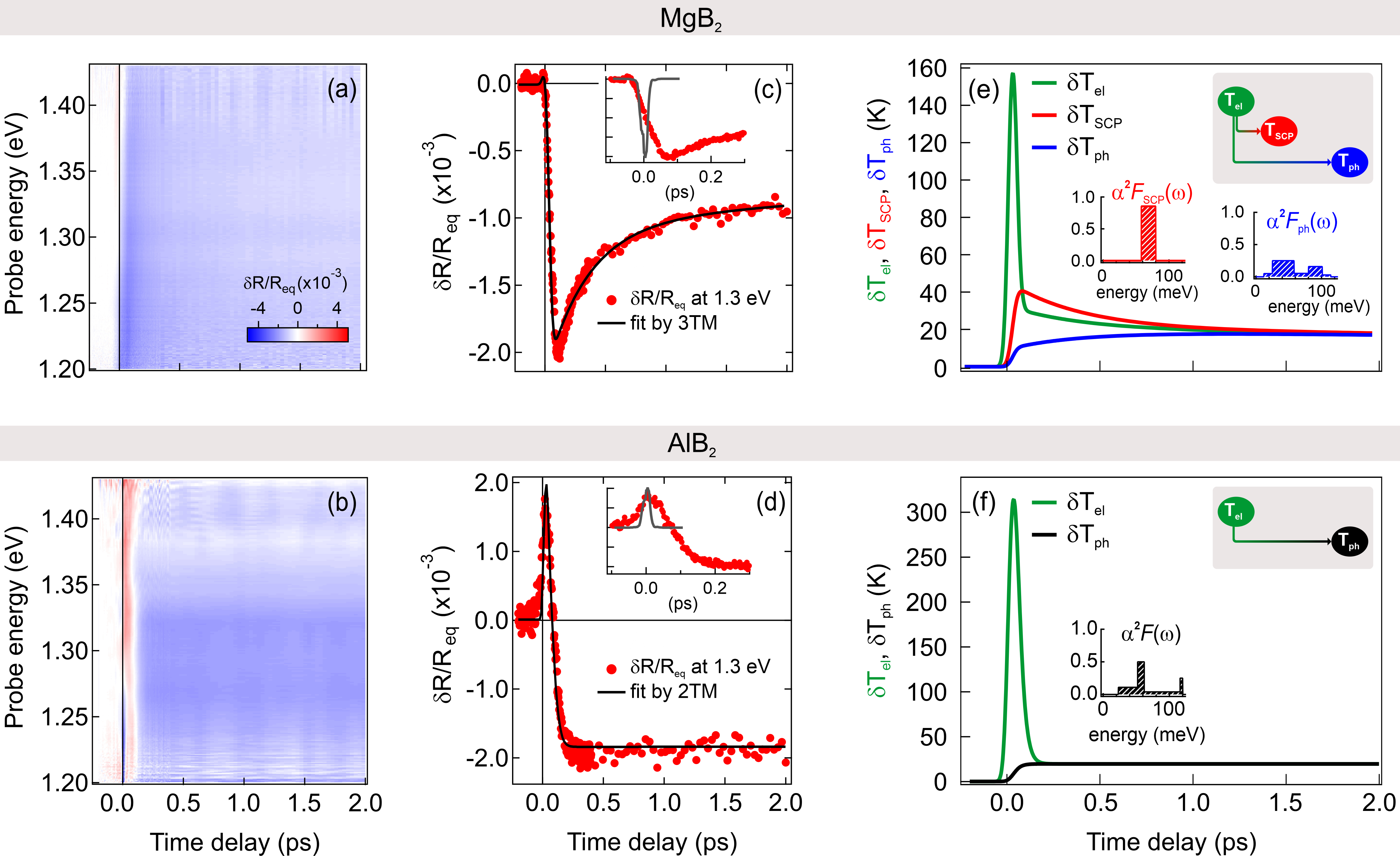}
\caption{$\delta R(t)/R_{eq}$ intensity maps \textcolor{black}{at incident fluence of 0.36 mJ cm$^{-2}$} of (a) \Mgb\, and (b) AlB$_2$ as a function of pump-probe time delay and probe photon energy. (c, d) Time-resolved line profiles at 1.3~eV (red markers) of the maps reported in (a) and (b), respectively, and relevant fits (black line) via effective-temperature modeling. The insets show the same line profiles up to 0.3~ps overlaid with the cross-correlation signal between pump and probe pulses (gray line), as measured through frequency optical gating technique~\cite{DalConte2015}. (e) Sketch of the temporal evolution of $\textcolor{black}{\delta} T_{el}$, $\textcolor{black}{\delta} T_{SCP}$ and $\textcolor{black}{\delta} T_{ph}$ in \Mgb\, after photoexcitation. The two dynamics are controlled by $\alpha^2\bar{F}_{SCP}$ and $\alpha^2\bar{F}_{ph}$ (red and blue histogram functions). (f) Sketch of the temporal evolution of $\textcolor{black}{\delta} T_{el}$ and $\textcolor{black}{\delta} T_{ph}$, and relevant $\alpha^2\bar{F}_{ph}$ (black histogram function) in AlB$_2$.}
\label{Fig:Fig3}
\end{figure*}

We performed broadband tr-OS of MgB$_2$ and AlB$_2$ in the probe energy range between 1.20 eV and 1.43 eV, which is lower than both the plasma edge (2 eV) and the MgB$_2$~$\sigma$-$\pi$ interband transitions centered at 2.78 eV~\cite{Baldini2017}, where anomalous blue-shifting dynamics of the plasmon has been observed~\cite{Baldini2017}. Therefore, in the probed energy range, a positive transient reflectivity variation would primarily trace the temporal evolution of the electronic temperature $T_{el}(t)$, whereas a negative one the effective phononic temperature, $T_{ph}(t)$. The measurements were performed at room temperature under photoexcitation with 2.1~eV pump pulses. The reflected probe beam was spectrally dispersed by a grating and detected by an array of photodetectors. Both pump and probe pulses were compressed below 20 fs as measured from the full width at half maximum of the cross-correlation signal by the frequency-resolved optical gating method, shown in the inset of Figure \ref{Fig:Fig3} (c). The time zero $t_0$, i.e. the time of pump-probe pulses' overlap, is defined as the maximum of the cross-correlation signal. The measurements were performed within a linear regime of the optical response (see Section II of the Supplemental Material for details~\cite{Supplementary}). \textcolor{black}{A detailed discussion of the effect of pump fluence variation on the transient optical response and the effective temperature models can be found in Section V of the Supplemental Material~\cite{Supplementary} (see also references~\cite{Ansari2020,Putti2002,Wang2015,CAPPELLUTI2022,Sentef2013,Ali1993,DellaValle2012,Obergfell2020,Waldecker2016} therein)}. Further details on the experimental setup can be found in Ref.~\citenum{DalConte2015}.

Figures \ref{Fig:Fig3} (a) and (b) report the transient reflectivity variation $\delta R(\omega,t)/R_{eq}$ of \Mgb\, and AlB$_2$, respectively, as a function of the pump-probe time delay (bottom axis) and the probe photon energy (left axis). The transient reflectivity variation of \Mgb\, is characterized by a negative signal in the entire spectral range and at any time delay. This points toward an almost instantaneous thermalization of electrons with at least a subset of SCPs, which implies that the reflectivity variation is dominated by the negative contribution due to the increase of electron-phonon scattering rate. 
Accordingly, we can assume $\delta R(t)/R_{eq}$ at 1.3 eV (red markers in Figure \ref{Fig:Fig3} (c)) maps the rise of $T_{\mathrm{SCP}}$ within $\tau_{\text{build-up}}\simeq$110 fs, before a two-step recovery dynamics takes place on time scales on the order of 400~fs and several ps (i.e., longer than the temporal window explored in the experiment). \textcolor{black}{For comparison with the literature, we mention that a $\delta R(t)/R_{eq}$ signal with positive sign was reported at ca. 1.5 eV~(see Supplementary information in Ref. \onlinecite{DalConte2015}. However, this signal is present only at time zero and is time-resolution limited. For these reasons, its origin can be attributed to a coherent artifact known as pump perturbed free induction decay~\cite{Brito1988} and not to any thermalization process. In view of that, in the present work, we avoided this spectral range and carried out our analysis where the positive coherent artifact was negligible compared to the negative pump-probe signal. Furthermore, previous transient reflectivity of MgB$_2$ thin films reported an initial positive dynamics at 1.5 eV, which decays on a $\sim$150 fs time scale due to SCP emission, eventually evolving towards a negative signal linked to electron-phonon scattering processes~\cite{Demsar2003,Xu2003}. We attribute the different sign of the initial transient reflectivity in thin films to either a lower frequency of the plasma edge, beyond which the reflectivity change associated to an increased scattering rate reverses sign~\cite{Giannetti2011,DalConte2012,Cilento2014,DalConte2015}, or a larger $T_{el}$ increase, which outweighs the opposing negative contribution arising from rapid SCP emission.}

The response of AlB$_2$ is both qualitatively and quantitatively different: the positive $\delta R(t)/R_{eq}$ right after time zero and at any probe energy reveals an instantaneous increase of the electron temperature. Then, the signal shows a decay with a single time constant of $\tau_{1}$ = 74 $\pm$ 8~fs leading to a sign change and a quasi-constant negative intensity that persists for several picoseconds (see $\delta R(t)/R_{eq}$ at 1.3 eV reported in Figure \ref{Fig:Fig3} (d)). This dynamics evidences an isotropic coupling of 'hot' electrons with the entire phonon bath and the heating of the lattice on the sub-ps time scale.

We now make use of the effective temperature model to reproduce the tr-OS data of \Mgb\, and AlB$_2$ and extract the relevant EPC strengths. We first consider the case of \Mgb. In this material, the strong and selective EPC is expected to result in a relaxation dynamics occurring on two time scales: the faster one is determined by scattering with the subset of E$_{2g}$ SCPs; the slower one by energy transfer from electrons to the weakly-coupled phonons. These two-steps dynamics can be described by a set of coupled differential equations for the three-temperature model (3TM) :~\cite{Allen1987,Perfetti2007}: 

\begin{align*}\label{Eqn:3Temp}
\frac{\partial T_{el}}{\partial t}&=\frac{g\left(\alpha^2\bar{F}_{SCP},T_{el},T_{SCP}\right)}{\gamma_{el}T_{el}} + \\&~~~~\frac{g\left(\alpha^2\bar{F}_{ph},T_{el},T_{ph}\right)}{\gamma_{el}T_{el}}+\frac{P(t)}{\gamma_{el}T_{el}} \\
\frac{\partial T_{SCP}}{\partial t}&=-\frac{g\left(\alpha^2\bar{F}_{SCP},T_{el},T_{SCP}\right)}{C_{SCP}} \\
\frac{\partial T_{ph}}{\partial t}&=-\frac{g\left(\alpha^2\bar{F}_{ph},T_{el},T_{ph}\right)}{C_{ph}}-\frac{T_{ph}-T_{0}}{\tau_{anh}} \\
\end{align*}
where the functionals
$g(\alpha^2\bar{F}_{i},T_{el},T_{i})$ are given by Eq. \ref{G} (with $i = SCP,ph$) and control the energy transfer rate between the 'hot' electron distribution and the two phonon populations at $T_{SCP}$ and $T_{ph}$, respectively. $\alpha^2\bar{F}_{SCP, ph}$ are the couplings with the two subsets of phonon modes that linearly contribute to the total glue function 
previously determined from the equilibrium OS data (see Section \ref{sec:OS}). \textcolor{black}{We point out that interband and interband coupling terms are not explicitly included in the 3TM, as the momentum-integrated transient reflectivity signal does not allow for differentiation between the two electron scattering processes.} The source term $P(t)$ in the first differential equation is a Gaussian function accounting for the pump pulse fluence and the experimental temporal resolution (the pump-pulse cross correlation is shown as gray lines in the insets of Fig. \ref{Fig:Fig3}(c) and (d)). The term $\gamma_{el}T_{el}$ expresses the electronic specific heat capacity, where the Sommerfeld constant $\gamma_{el}$ is taken as 1.5$\,$10$^{-4}\frac{J}{K^{2}cm^{3}}$~\cite{Junod2001}. $C_{SCP}$ and $C_{ph}$ are the specific heat capacities of the SCPs and the rest of the phonons. With SCPs being a fraction $f$ of the total number of phonon modes, we assume $C_{SCP} = fC_{lat}$, where $C_{lat}$ = 3$\frac{J}{Kcm^{3}}$~\cite{Junod2001} is the total specific heat capacity of the lattice. Finally, the third differential equation contains a term that accounts for the anharmonic phonon decay\textcolor{black}{~\cite{Mialitsin2007}} due to phonon-phonon scattering on a time scale $\tau_{anh}$ much longer than the temporal window explored in the experiments. Because we expect the major part of $\alpha^2\bar{F}(\omega)$ at $\sim$70 meV to be determined by the SCP modes, we assume the following expression to hold: $\alpha^2\bar{F}_{SCP} = p\,\alpha^2\bar{F}(70~\text{meV})$, with $0<p<1$. It then follows that 
$\alpha^2\bar{F}_{ph} = \alpha^2\bar{F}(\omega)-p\,\alpha^2\bar{F}(70~\text{meV})$. 

Based on these assumptions, Figure \ref{Fig:Fig3}(c) shows the best fit (black line) to $\delta R(t)/R_{eq}$ of \Mgb\, at 1.3~eV obtained as a linear combination of the contributions given by the temporal evolution of $T_{el}$, $T_{SCP}$ and $T_{ph}$. 
Those are obtained by solving the effective 3TM explained above, where $p$ and $f$ are fitting parameters. We find that the major contribution to $\delta R(t)/R_{eq}$ stems from the temporal evolution of $T_{\mathrm{SCP}}$ that is associated with the broadening of the Drude peak due to increased electron-phonon scattering rate. In particular, the latter exceeds by a factor of 10 the contribution due to the evolution of the electronic temperature, explaining the absence of any positive contribution to the transient reflectivity signal. Moreover, the SCP participating in the thermalization dynamics constitutes a rather small fraction $f = 0.28\,\pm0.03$ of the total phonon modes\textcolor{black}{, resulting in $C_{SCP} = 0.84 \frac{J}{Kcm^{3}}$}. However, their contribution to the coupling strength is $p = 0.94\,\pm0.06$, which is a significant portion of the $\alpha^2\bar{F}$ spectrum at 70~meV. The fit results allow us to decompose the $\alpha^2\bar{F}$ function into a term due to SCP and one due to the remaining lattice modes, and to calculate the respective contribution to $\lambda$ using Eq. (2). In particular, we obtain $\lambda_{SCP}$ = 0.56, which corresponds to approximately 0.48 times the total EPC constant of MgB$_2$, in remarkable agreement with the theoretically predicted ratio of 0.43~\cite{Novko2020}. Finally, the coupling to the other lattice modes is $\lambda_{ph}$ = 0.61. Figure \ref{Fig:Fig3}(e) displays the evolution of \textcolor{black}{changes in} $T_{el}$, $T_{SCP}$ and $T_{ph}$ in \Mgb, together with the associated $\alpha^2\bar{F}_{SCP,ph}$. The rapid increase of $T_{\mathrm{SCP}}$ reflects the decrease of $T_{el}$ due to energy transfer to the SCPs, and accounts for both the absence of a positive signal and the build-up of a negative-intensity $\delta R(t)/R_{eq}$ (cf. $\tau_{\text{build-up}}$). The thermalization with the rest of the phonon bath described by $\textcolor{black}{\delta}T_{ph}$ reproduces the subsequent two-steps dynamics (cf. $\tau_{\text{fast}}$ and $\tau_{\text{slow}}$). All-in-all, tr-OS of \Mgb\,  highlights the existence of two different subsets of phonon modes that are directly coupled to electrons with different EPC strengths. 
The SCP subset enables rapid energy transfer from electrons to the $E_{2g}$ phonons, acting as energy reservoir that prevents from heating of the whole lattice. $T_{ph}$ due to the other weakly-coupled phonon modes refrains from a rapid increase and reaches a maximum only after approximately 1.5 ps, i.e. until a common temperature is reached among the electronic and phononic subsets.

We benchmark these results against AlB$_2$ (Fig.\ref{Fig:Fig2}(d)), which does not host strong EPC with specific phonon modes. In this case, $\delta R(t)/R_{eq}$ at 1.3~eV is well reproduced by a two-temperature model (2TM, black line) that includes $T_{el}$ and $T_{ph}$. Specifically, the initial $\delta R(t)/R_{eq}$ variation originates from the increase of $T_{el}$ whose evolution is decoupled from that of $T_{ph}$. The delayed negative signal is associated to the increase of $T_{ph}(t)$, whose single-step build-up dynamics is determined by the weak and isotropic coupling of electrons with the lattice, ($\lambda$(AlB$_2$) = 0.42). Eventually, the energy stored in the phononic bath is expected to relax due to phonon anharmonic decay on the time scale $\tau_{anh}$ much longer than the explored energy window. \textcolor{black}{We point out that the fact that $T_{el}$ and $T_{ph}$ reach the same value on a faster time scale as compared to \Mgb\, does not imply stronger EPC which is, in fact, much smaller~\cite{Bohnen2001}. Instead, \textcolor{black}{
we recall that the electron-phonon energy transfer rate depends not only on the coupling strength, but also on the specific heat of the receiving subsystem \textcolor{black}{that we assume to be proportional to the number of phonons involved in the thermalization dynamics}. Thus, in AlB$_2$, } the energy transfer from the photoexcited electrons to the lattice \textcolor{black}{ occurs faster since it} involves equally the entire phonon spectrum. This is, in effect, at odds with the predicted behavior in \Mgb, where lattice heating is delayed due to the formation of a \textcolor{black}{rather small ($f = 0.28$)} non-thermal population of hot $E_{2g}$ phonons~\cite{Novko2020}.}

\section{Discussion and conclusions}
In this work, we have demonstrated that tr-OS enables directly accessing the energy exchange dynamics between electrons and the subset of more strongly coupled phonon modes. In the case of the BCS superconductor MgB$_2$, we have shown that the strong and selective coupling with the E$_{2g}$ B-B stretching phonon modes at $\sim$70 meV gives rise to an additional relaxation dynamics, which is absent in the non-superconducting isostructural compound AlB$_2$. The coupling strength with these SCPs, known to determine the $T_c$ of the superconducing phase transition, is $\lambda_{SCP}$(MgB$_2$) = 0.56, thus approximately half of the total EPC constant of \Mgb. This ratio is in agreement with recent estimations by first-principles quantum field theory calculations \cite{Novko2020}.  \textcolor{black}{We then use the obtained $\lambda_{SCP}$(MgB$_2$) to estimate $T_c$ using Eq. \ref{eq:Tc}. In the limit of $\mu^* = 0$, we calculate the upper bound of $T_c$ = 35~K, thus very close to the actual $T_c$. The accounting of the pseudopotential $\mu^*$ is not trivial, as the literature reports values from 0.05~\cite{Bohnen2001} up to 1.151\cite{Moon2004}. For a mean value of $\mu^* = 0.1$, we obtain $T_c$ = 17~K, which is still more than 40\% of the actual value. These findings corroborate theory predictions that the strong and selective EPC with $E_{2g}$ phonons is the driving force for the superconducting pairing in \Mgb\, as well as a necessary condition for the superconductivity to appear at temperatures as high as 40 K~\cite{Yildirim2001, Kong2001, Bohnen2001}.} Eventually, the transient decoupling between the SCP distribution and the other phonon modes opens up the possibility of selectively controlling the nonequilibrium population of specific SCP modes \cite{Baldini2017,Novko2020}, thus transiently modifying the relevant electronic properties, including superconductivity. 

Despite lacking the characteristic momentum resolution of ARPES or the wide frequency resolution of equilibrium OS, nonequilibrium spectroscopy provides a direct and quantitative estimation of the selective coupling with specific boson excitations and allows the disentanglement of the different electron-boson coupling channels $i$ that contribute to the total electron-boson coupling function, i.e. $\alpha^2\bar{F}$=$\sum_i\alpha^2\bar{F}_i$. This methodology can be extended to unconventional superconductors, in which charge, spin and orbital excitations emerge from strong electronic correlations \cite{Giannetti2017}. Moreover, quantitative access to EPC in anisotropically coupled electronic systems will be crucial for a robust control of the energy transfer mechanisms on the ultrafast time scales, including the manipulation of 'hot' phonons physics in order to limit overall heating of photoexcited systems and hence sustain quantum-coherent states~\cite{Cappelluti2022_2}.\\

The work at UBC was undertaken thanks in part to funding from the Max Planck-UBC-UTokyo Centre for Quantum Materials and the Canada First Research Excellence Fund, Quantum Materials and Future Technologies Program. This project is also funded by the Canada Foundation for Innovation (CFI); the British Columbia Knowledge Development Fund (BCKDF); the Natural Sciences and Engineering Research Council of Canada (NSERC); the Department of National Defence (DND); the Fonds de recherche du Qu\'{e}bec --- Nature et Technologies (FRQNT); the Minist\`{e}re de l'\'{E}conomie, de l'Innovation et de l'\'{E}nergie --- Qu\'{e}bec; the Canada Research Chair Program (A.D. and F.B.); and the CIFAR Quantum Materials Program (A.D.). C.G.   acknowledges financial support from MIUR
through the PRIN 2017 (Prot. 20172H2SC4 005) and PRIN 2020 (Prot.
2020JLZ52N 003) programs and from the European Union - Next Generation EU, Missione 4 Componente 2, CUP J53D23001380008. C.G. and A.D. received support for collaborative research from the Peter Wall Institute for Advances Studies. G.C. and S.D.C. acknowledge support from the European Union’s Next Generation EU Programme with the I-PHOQS Infrastructure
423 [IR0000016, ID D2B8D520, CUP B53C22001750006] “Integrated infrastructure initiative in Photonic and Quantum Sciences”. Work done at Ames National Laboratory (PCC - crystal growth) was supported by the U.S. Department of Energy, Office of Basic Energy Science, Division of Materials Sciences and Engineering.  Ames Laboratory is operated for the U.S. Department of Energy by Iowa State University under Contract No. DE-AC02-07CH11358.

\appendix*
\section{Sample growth}
Single crystals of MgB$_2$ were grown under high pressure using the cubic anvil press~\cite{Karpinski2007}. A BN crucible with an inner diameter of 8~mm and a length of 9~mm was filled with a mixture of Mg, B, and BN in a 10:12:1 ratio. The typical growth procedure was the following: increasing the pressure to 30~kbar at room temperature, raising the temperature to a maximum value of 1960°C in 1~h, holding it there for 1 h, and then lowering the temperature and pressure in 1.5~h. Magnetic and x-ray diffraction measurements confirmed the high quality of the produced crystals with a superconducting transition at 38.7~K.

Single crystals of AlB$_2$ were grown out of a very Al-rich self flux~\cite{Canfield2020}.  Al and B were place into a fritted crucible set~\cite{Canfield2016} (sold as Canfield Crucible Sets, or CCS, by LSP ceramics)~\cite{CCS} in an atomic ratio of Al$_{0.945}$B$_{0.055}$.  The CCS was then sealed into an amorphous
silica tube that was evacuated and then back-filled with $\sim$ 1/6 ATM of high purity Ar.  The sealed ampoule was then place in a box furnace, heated to 1180°C over 15 hours, held at 1180°C for 10 hours, and then cooled to 710°C over 125 hours.  At 710°C the ampoule was removed from the furnace and placed in a centrifuge for decanting~\cite{Canfield2020}.  The excess Al liquid was removed from the thin AlB$_2$ plates.

\bibliography{MgB2}

\end{document}